\documentclass[showpacs,preprint]{revtex4-1}
\usepackage{amssymb}
\usepackage{amsmath}
\usepackage{graphicx}
\usepackage{bm}

\begin{document}

\title{Curved and diffuse interface effects on the nuclear surface tension}

\author{V.M.~Kolomietz, S.V.~Lukyanov and A.I.~Sanzhur}

\begin{abstract}
We redefine the surface tension coefficient for a nuclear Fermi-liquid drop
with a finite diffuse layer. Following Gibbs-Tolman concept, we introduce
the equimolar radius $R_{e}$\ of sharp surface droplet at which the surface
tension is applied and the\ radius of tension surface $R_{s}$ which provides
the minimum of the surface tension coefficient $\sigma$. This procedure
allows us to derive both the surface tension and the corresponding curvature
correction (Tolman length) correctly for the curved and diffuse interface.
We point out that the curvature correction depends significantly on the
finite diffuse interface. This fact is missed in traditional nuclear
considerations of curvature correction to the surface tension. We show that
Tolman's length $\xi $ is negative for nuclear Fermi-liquid drop. The value
of the Tolman length is only slightly sensitive to the Skyrme force
parametrization and equals $\xi =- 0.36$ fm.
\end{abstract}

     
\maketitle

\affiliation{Institute for Nuclear Research, 03680 Kiev, Ukraine}

\section{Introduction}

\label{intro} The binding energy of saturated many-particle systems like a
nucleus exhibits the hierarchy of expansion in powers of mass number $%
A^{1/3} $ which is associated with well-established experimental features
and plays an important role in the understanding of macroscopic properties
of nuclei \cite{BoMo}. In a simplest case, the surface energy $E_{\mathcal{S}%
}$ of such systems is given by the term of order $A^{2/3}$ in this
hierarchy. The structure of the surface energy $E_{\mathcal{S}}$ and the
corresponding surface tension coefficient $\sigma $ depend on the
interparticle interaction and the surface conditions. Moreover, the nucleus
is a two component, charged system with a finite diffuse layer. This fact
specifies a number of various peculiarities of the nuclear surface energy $%
E_{\mathcal{S}}$: dependency on the density profile function, contribution
to the surface symmetry energy, connection to the nuclear incompressibility,
etc. The additional refinements of $E_{\mathcal{S}}$ appear due to the
quantum effects arising from the smallness of nucleus. In particular, the
curved interface creates the curvature correction to $E_{\mathcal{S}}$ of
order $A^{1/3}$ and can play the appreciable role in small nuclei.

The curvature correction to the planar tension coefficient and the
corresponding Tolman length \cite{tolm49} can be estimated
phenomenologically using the polynomial, in powers of $A^{1/3}$, expansion
of mass formula \cite{Myer69,mysw98,Stoc1973,tyap71}. The importance of
curvature term for the evaluation of nuclear masses and fission barriers and
the interplay between different terms of the $A^{1/3}$-expansion due to the
value of the radius was shown in \cite{podu03}. However the influence of the
curved interface on the properties of small quantum systems is still poorly
studied because of the finite diffuse layer where particle density drops
down to the zero value. The presence of the finite diffuse layer in a small
drop creates two, at least, questions: (i) What is the actual radius of a
drop? (ii) What is the physical surface where the surface tension is
applied? Since the presence of the diffuse layer, different definitions for
the size of the drop are possible \cite{Myer1973} which all give the value
of the radius of spherical drop located within the diffuse layer.

Note that a small indefiniteness in a nuclear radius of the order of $\Delta
R\approx 0.5$\ fm, i.e., within the diffuse layer, leads to a shift of
surface energy of order $\Delta E_{\mathcal{S}}\approx 10^{2}$ $\mathrm{MeV}$
(for $^{208}$Pb \cite{BoMo}) which exceeds significantly both the shell \cite%
{stru67} and the curvature \cite{mysw98} corrections to the nuclear binding
energy, see also ref.~\cite{podu03}. We point out also even though the width
of diffuse layer is much less than the range of approximate uniformity of
the particle density, one still needs the strict definition of the drop size
because of the following reason. In contrast to the planar geometry, the
area $S$\ for the spherical (curved) surface will depend on the choice of
drop radius and this will affect the value of the surface tension $\sigma $\
derived from the surface energy.

Gibbs was the first who addressed the problem of the correct definition of
the radius and the surface of tension to a small drop with a diffuse
interface \cite{gibbs}. After him, Tolman drew attention \cite{tolm49} that
two different radii have to be introduced in this case: the equimolar radius 
$R_{e}$, which gives the actual size of the corresponding sharp-surface
droplet, and the\ radius of tension $R_{s}$,\ which derives, in particular,
the capillary pressure, see below in sect. \ref{sec:2}. Following Tolman,
see also ref. \cite{Wido82}, the surface tension $\sigma _{e}\equiv \sigma
(R_{e})$ approaches the planar limit $\sigma _{\infty }$ as 
\begin{equation}
\sigma (R_{e})=\sigma _{\infty }\left( 1-\frac{2\xi }{R_{e}}+\mathcal{O}%
(R_{e}^{-2})\right) \ ,  \label{sigmaeq}
\end{equation}%
where $\xi $ is the Tolman's length \cite{tolm49}. At the same time\ the
capillary pressure $P_{\mathrm{capil}}$, which is generated by the curved
surface and provides the equilibrium condition for the well-defined radius $%
R_{e}$, is derived by the\ radius of tension $R_{s}$ \cite{gibbs}%
\begin{equation}
P_{\mathrm{capil}}=\frac{2\sigma }{R_{s}}.  \label{capil}
\end{equation}

In general, the presence of the curved interface affects both the bulk and
the surface properties. The curvature correction $\Delta \sigma _{\mathrm{%
curv}}=-2\sigma _{\infty }\xi /R_{e}\sim A^{-1/3}$\ is usually negligible in
heavy nuclei. However, this correction can be important in some nuclear
processes. That the yield of fragments at the nuclear multifragmentation or
the probability of clasterization of nuclei from the freeze-out volume in
heavy ion collisions are derived by the statistical weight $W$ of the radius
fluctuations \cite{lali58} 
\[
W\propto e^{-\mathrm{const~}\sigma R^{2}/T}\ . 
\]%
In both above mentioned processes, the small nuclei occur necessarily and
the exponential dependence of statistical weight $W$ on the surface tension $%
\sigma $ should cause the sensitivity of both processes to the curvature
correction $\Delta \sigma _{\mathrm{curv}}$. The curvature correction can
also play appreciable role for the nuclear fission near the scission point
and for the nuclear fusion in the neck region. This aspect of large nuclear
deformations was not studied yet.

In nuclear physics, the curvature correction to the surface tension was
intensively investigated phenomenologically \cite{Myer69,mysw98,tyap71} as
well as within the quantum approaches \cite{FaSt1985}. Using the
Thomas-Fermi approximation, the dependence on curvature of the nuclear
surface energy was studied in ref. \cite{bera74} and the various terms of
the droplet model were derived from the Skyrme interaction in ref. \cite%
{brguho84}. In ref. \cite{gram83}, a general procedure restricted by the
Skyrme-type functional and the terms of the order $\hbar ^{2}$\ was applied
for calculation of the curvature energy and in ref. \cite{stbanisi88} the
curvature-energy was studied by use of semiclassical mean-field approaches
with including higher-order terms. In ref. \cite{far85} a closed expression
for the nuclear curvature energy and its expansion in series of volume terms
and surface moments was deduced in a soluble model. For the Fermi gas model
in an external Woods-Saxon potential the curvature energy was calculated in
ref. \cite{duschvi93}. In ref. \cite{podu03}, the nuclear liquid-drop model
containing the first and/or the second order curvature terms was revised and
it was reproduced with a reasonable precision the nuclear binding energies
and the fission-barriers based on large amount of up-to-date experimental
data. A special care was given to the kinetic energy operator and the bulk
density oscillations within the quantum-mechanical approach \cite{FaSt1985}.
It was shown that, in contrast to the semiclassical approaches, to obtain
the reasonable value for the curvature energy the particle and energy
densities should be averaged in a special way.

In present paper, we suggest the microscopic analysis of the curvature
correction to the surface tension of a small drop with a finite diffuse
layer. We follow the ideology of the extended Thomas-Fermi approximation
(ETFA) with effective Skyrme-like forces combining the ETFA and the direct
variational method. In our consideration, the proton and neutron densities $%
\rho _{p}(\mathbf{r})$ and $\rho _{n}(\mathbf{r})$ are generated by the
diffuse-layer profile functions which are eliminated by the requirement that
the energy of the nucleus be stationary with respect to variations of these
profiles. In order to formulate proper definition for the drop radius, we
will use the concept of dividing surface, originally introduced by Gibbs 
\cite{gibbs}. Following Gibbs, we will introduce the superficial (surface)
density as the difference \ (per unit area of dividing surface) between
actual number of particles $A$\ and the number of particles $A_{\mathcal{V}}$%
\ which drop would contain if the particle density retained uniform.

This paper is organized as follows. In sect.~\ref{sec:2} we give the
thermodynamical derivation of the surface tension for a finite system. The
Tolman length is derived in sect.~\ref{sec:3}. Numerical results and
conclusions are summarized in sects.~\ref{sec:4} and \ref{concl}. The
connection of Gibbs-Tolman approach to the droplet model is given in sect.~%
\ref{droplet}.

\section{Equimolar surface}

\label{sec:2} We will calculate the dependence of surface tension
coefficient on the position of the dividing surface in a small Fermi-liquid
drop with a finite diffuse layer similarly to procedure described in refs.~%
\cite{Wido82,Nijm92}. The goal of calculations is to determine the position
of the equimolar surface, the dependence of surface tension on the bulk
density and the sensitivity of the curvature correction (Tolman length $\xi $%
) to the parametrization of the effective nuclear forces.

We consider the uncharged % two-component
symmetric ($N=Z$) droplet having number of particles
$A=N+Z$, chemical potential $\lambda $ and free energy $F$. Note that the
thermodynamical consideration is most adequate here because of the finite
diffuse interface in a cold nucleus is similar to the vapor environment in a
classical liquid drop. In order to formulate proper definition for the drop
radius, we will use the concept of dividing surface of radius $R$,
originally introduced by Gibbs \cite{gibbs}. Following refs.~\cite%
{Wido82,gibbs}, we introduce the formal (arbitrary but close to the
interface) dividing surface of radius $R$, the corresponding volume $%
\mathcal{V}=4\pi R^{3}/3$ and the surface area $\mathcal{S}=4\pi R^{2}$. The
droplet free energy $F$ will be then split between volume, $F_{\mathcal{V}},$%
\ and surface, $F_{\mathcal{S}}$, parts%
\begin{equation}
F=F_{\mathcal{V}}+F_{\mathcal{S}}\ ,  \label{esplit}
\end{equation}%
where 
\begin{equation}
F_{\mathcal{V}}=\left( -P+\lambda \varrho _{\mathcal{V}}\right) \mathcal{V}\
,\ \ F_{\mathcal{S}}=\left( \sigma +\lambda \varrho _{\mathcal{S}}\right) 
\mathcal{S}\ .  \label{evolsurf}
\end{equation}%
Here, $P=P(\lambda )$ is the pressure of nuclear matter achieved at some
volume particle density $\varrho _{\mathcal{V}}=A_{\mathcal{V}}/\mathcal{V}$
and $\varrho _{\mathcal{S}}=A_{\mathcal{S}}/\mathcal{S}$ is the surface
density, where $A_{\mathcal{V}}$ and $A_{\mathcal{S}}$ are the volume and
the surface particle number, respectively. The actual particle number is
given by 
\begin{equation}
A=A_{\mathcal{V}}+A_{\mathcal{S}}=\varrho _{\mathcal{V}}\mathcal{V}+\varrho
_{\mathcal{S}}\mathcal{S}\ .  \label{asplit}
\end{equation}%
The use of eqs.~(\ref{esplit}) -- (\ref{asplit}) gives the following
relation for the surface tension 
\begin{equation}
\sigma =\frac{F-\lambda A}{\mathcal{S}}+\frac{P\mathcal{V}}{\mathcal{S}}=%
\frac{\Omega -\Omega _{\mathcal{V}}}{\mathcal{S}}\ ,  \label{sigma}
\end{equation}%
where symbol $\Omega =F-\lambda A$ stands for the grand potential and $%
\Omega _{\mathcal{V}}=-P\mathcal{V}$. To reveal a $R$-dependence of surface
tension $\sigma $, it is convenient to introduce the grand potential per
particle $\omega =F/A-\lambda $ for the actual droplet and $\omega _{%
\mathcal{V}}=F_{\mathcal{V}}/A_{\mathcal{V}}-\lambda =-P/\varrho _{\mathcal{V%
}}$ for the volume part. Then the surface tension is written as 
\begin{equation}
\sigma \left[ R\right] =\frac{\omega A}{4\pi R^{2}}-\frac{1}{3}\omega _{%
\mathcal{V}}\varrho _{\mathcal{V}}R\ .  \label{sigmaR}
\end{equation}%
Here, the square brackets denote a dependence of the observable $F$, $%
\lambda $, $P$ etc. on the dividing surface radius $R$\ which is different
than the dependence on the physical size of a droplet \cite{Nijm92}. Using
eq.~(\ref{sigma}), the capillary pressure $P$ reads 
\begin{equation}
P=3\frac{\sigma \left[ R\right] }{R}-3\frac{F-\lambda A}{4\pi R^{3}}\ .
\label{p1}
\end{equation}%
Taking the derivative from eq.~(\ref{p1}) with respect to the formal
dividing radius $R$ and using the fact that the observable $F,$ $\lambda $
and $P$ are $R$-independent (changing dividing radius $R$ we keep the
particle density invariable), one can rewrite eq.~(\ref{p1}) as 
\begin{equation}
P=2\frac{\sigma \left[ R\right] }{R}+\frac{\partial }{\partial R}\,\sigma %
\left[ R\right] \ ,  \label{p2}
\end{equation}%
which is the generalized Laplace equation.

The choice of the dividing radius $R$ is arbitrary, the only condition is to
keep the same chemical potential $\lambda $. So, the formal value of surface
density $\varrho _{\mathcal{S}}$ can be positive or negative depending on $R$%
. From eq.~(\ref{asplit}) one finds 
\begin{equation}
\varrho _{\mathcal{S}}[R]=\frac{A}{4\pi R^{2}}- \frac{1}{3}\varrho_{\mathcal{%
V}}R\ .  \label{gammaR}
\end{equation}

We have performed the numerical calculations using Skyrme type of the
effective nucleon-nucleon interaction. The energy and the chemical potential
for actual droplet have been calculated using direct variational method
within the extended Thomas-Fermi approximation \cite{KoSa08}. Assuming the
leptodermous condition, the total energy takes the following form of $A,X$%
-expansion \ 
\begin{equation}
F/A\equiv e_{A}=e_{0}(A)+e_{1}(A)X+e_{2}(A)X^{2}\ ,  \label{e1}
\end{equation}%
where $X$ is the isotopic asymmetry parameter $X=(N-Z)/A$ and 
\begin{equation}
e_{i}(A)=c_{i,0}+c_{i,1}A^{-1/3}+c_{i,2}A^{-2/3}\ .  \label{e2}
\end{equation}%
An explicit form of coefficients $c_{i,j}$ for the Skyrme forces is given in
ref.~\cite{KoSa08}. Using the trial profile function for the neutron $\rho
_{n}(r)$ and proton $\rho _{p}(r)$ densities and performing the direct
variational procedure, we can evaluate the equilibrium particle densities $%
\rho _{\pm }(r)=\rho _{n}(r)\pm \rho _{p}(r)$, equilibrium bulk density $%
\rho _{\pm ,0}=\lim\limits_{r\rightarrow 0}{\rho _{\pm }(r)}$, total free
energy per particle $F/A$ and chemical potentials $\lambda _{n}$\ and $%
\lambda _{p}$, %chemical potential $\lambda $,
see ref. \cite{KoSa08} for details. %Below we will use
%simplified consideration for the case $X=0$\ with no Coulomb interaction.
>From now on, we will consider symmetric and uncharged nuclei with $\lambda
_{n}=\lambda _{p}=\lambda $\ and $X=0$.

The volume part of free energy $F_{\mathcal{V}}/A_{\mathcal{V}}$ is
associated with coefficient $c_{0,0}$ of ref.~\cite{KoSa08} 
\[
F_{\mathcal{V}}/A_{\mathcal{V}}=c_{0,0},\ \ \ \ \ \ c_{0,0}=\frac{\hbar ^{2}%
}{2m}\alpha \rho _{+,0}^{2/3}+\frac{3t_{0}}{8}\rho _{+,0}+\frac{t_{3}}{16}%
\rho _{+,0}^{\nu +1}+ 
\]%
\begin{equation}
\frac{\alpha }{16}\left[ 3t_{1}+t_{2}(5+4x_{2})\right] \rho _{+,0}^{5/3}\ ,
\label{c00}
\end{equation}%
where $\alpha ={(3/5)}\,(3\,\pi ^{2}/2)^{2/3}$ and $t_{i}$, $x_{2}$ and $\nu 
$ are the Skyrme force parameters. Using the evaluated chemical potential $%
\lambda $, we fix the particle density $\varrho _{\mathcal{V}}=\varrho _{%
\mathcal{V}}(\lambda )$ from the condition 
\begin{equation}
\left. \frac{\partial F_{\mathcal{V}}}{\partial A_{\mathcal{V}}}\right\vert
_{\mathcal{V}}=\left. \frac{\partial }{\partial \rho _{+,0}}(\rho
_{+,0}c_{0,0})\right\vert _{\rho _{+,0}=\varrho _{\mathcal{V}}}=\lambda \ .
\label{volpart}
\end{equation}%
For an arbitrary dividing radius $R$ we evaluate then the volume particle
number $A_{\mathcal{V}}=4\pi R^{3}\varrho _{\mathcal{V}}/3$ and the volume
part of free energy $F_{\mathcal{V}}/A_{\mathcal{V}}$. Finally, evaluating
the surface parts $A_{\mathcal{S}}=A-A_{\mathcal{V}}$ and $F_{\mathcal{S}%
}=F-F_{\mathcal{V}}$, we obtain the surface tension coefficient $\sigma %
\left[ R\right] $ for an arbitrary radius $R$ of dividing surface.

\begin{figure}
\includegraphics[width=0.95\columnwidth]{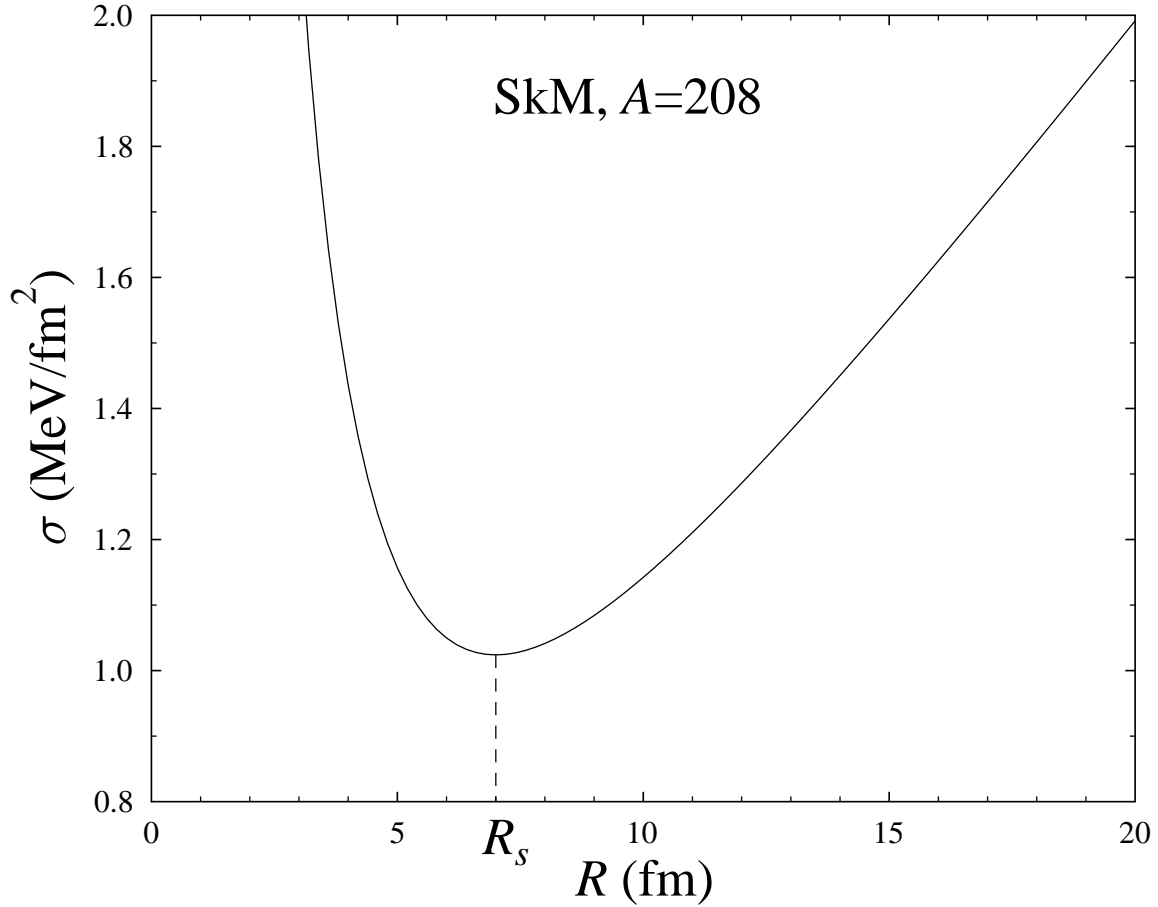}
\caption{Surface tension $\protect\sigma $\ as a function of the dividing
radius $R$ for $A=208$. Calculation was performed using $SkM$ force. $R_{s}$
denotes the dividing radius where $\protect\sigma$ approaches the minimum
value.}
\label{fig1}
\end{figure}

The dependence of the surface tension $\sigma \left[ R\right] $ on the
location of the dividing surface for $A=208$ is shown in \figurename\ \ref{fig1}. 
As seen from \figurename\ \ref{fig1}, function $\sigma \left[ R\right] $ has minimum
at radius $R=R_{s}$ (radius of surface of tension \cite{Wido82}) which
usually does not coincide with the equimolar radius $R_{e}$. The radius $%
R_{s}$ denotes the location within the interface. Note that for $R=R_{s}$
the capillary pressure of eq.~(\ref{p2}) satisfies the classical Laplace
relation 
\begin{equation}
P=2\left. \frac{\sigma \left[ R\right] }{R}\right\vert _{R=R_{s}}\ .
\label{p3}
\end{equation}

\section{Surface tension and Tolman length}

\label{sec:3} 
\begin{figure}
\includegraphics[width=0.95\columnwidth]{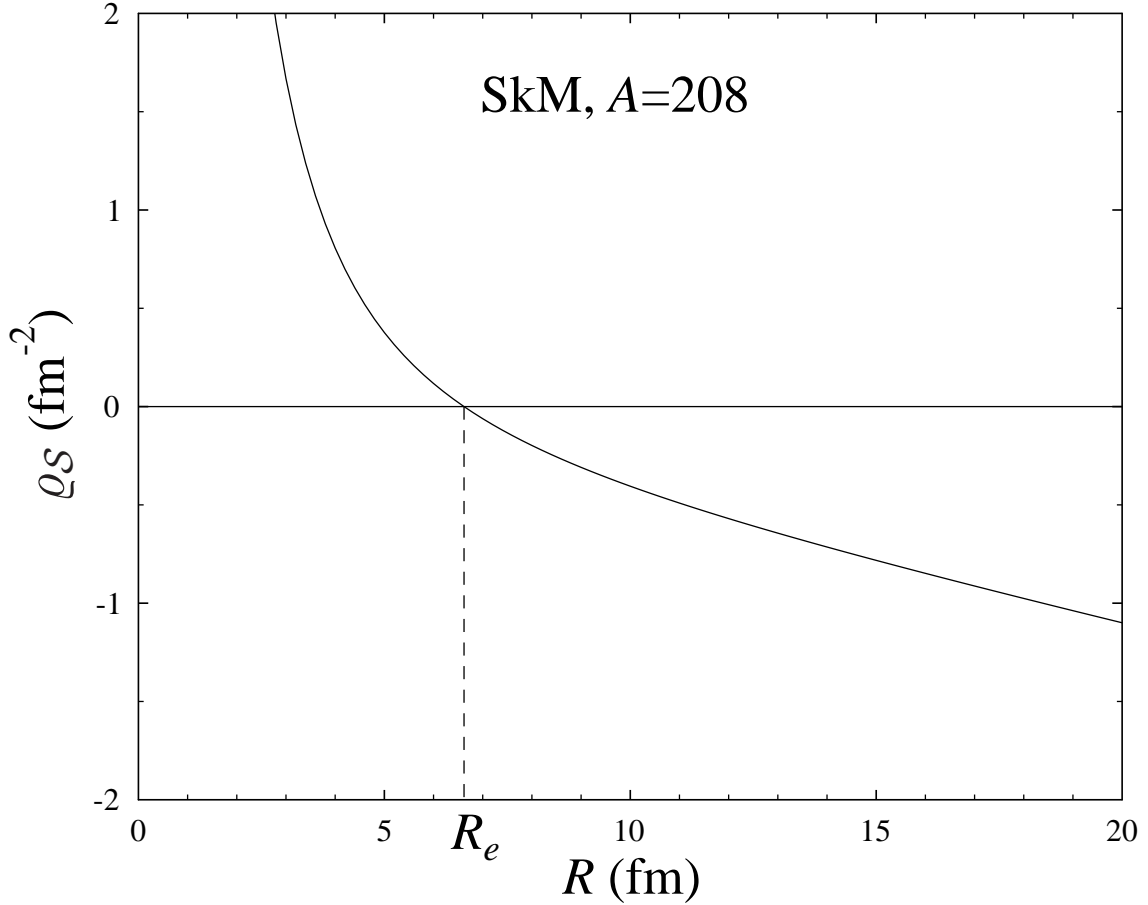}
\caption{Surface particle density $\protect\varrho_{\mathcal{S}}$ versus
dividing radius $R$ for $A=208$. Calculation was performed using SkM force. $%
R_{e}$ denotes the equimolar radius where $\protect\varrho _{\mathcal{S}}$
becomes zero.}
\label{fig2}
\end{figure}

In \figurename\ \ref{fig2} we present the calculation of the surface
particle density $\varrho _{\mathcal{S}}[R]$. Note that, in general, the
surface free energy $F_{\mathcal{S}}$\ includes both contributions from the
surface tension $\sigma $\ itself and from the bulk binding energy of $A_{%
\mathcal{S}}$\ particles within the surface layer. The equimolar surface and
the actual physical radius $R_{e}$\ of the droplet is derived by the
condition $\varrho _{\mathcal{S}}[R_{e}]=0$ \cite{tolm49,Wido82,Nijm92},
i.e., the contribution from the bulk binding energy should be excluded from
the surface free energy $F_{\mathcal{S}}$. The corresponding radius is
marked in \figurename\ \ref{fig2}. Equimolar dividing radius $R_{e}$ defines the
physical size of the sharp surface droplet and the surface at which the
surface tension is applied. 

\begin{figure}
\includegraphics[width=0.95\columnwidth]{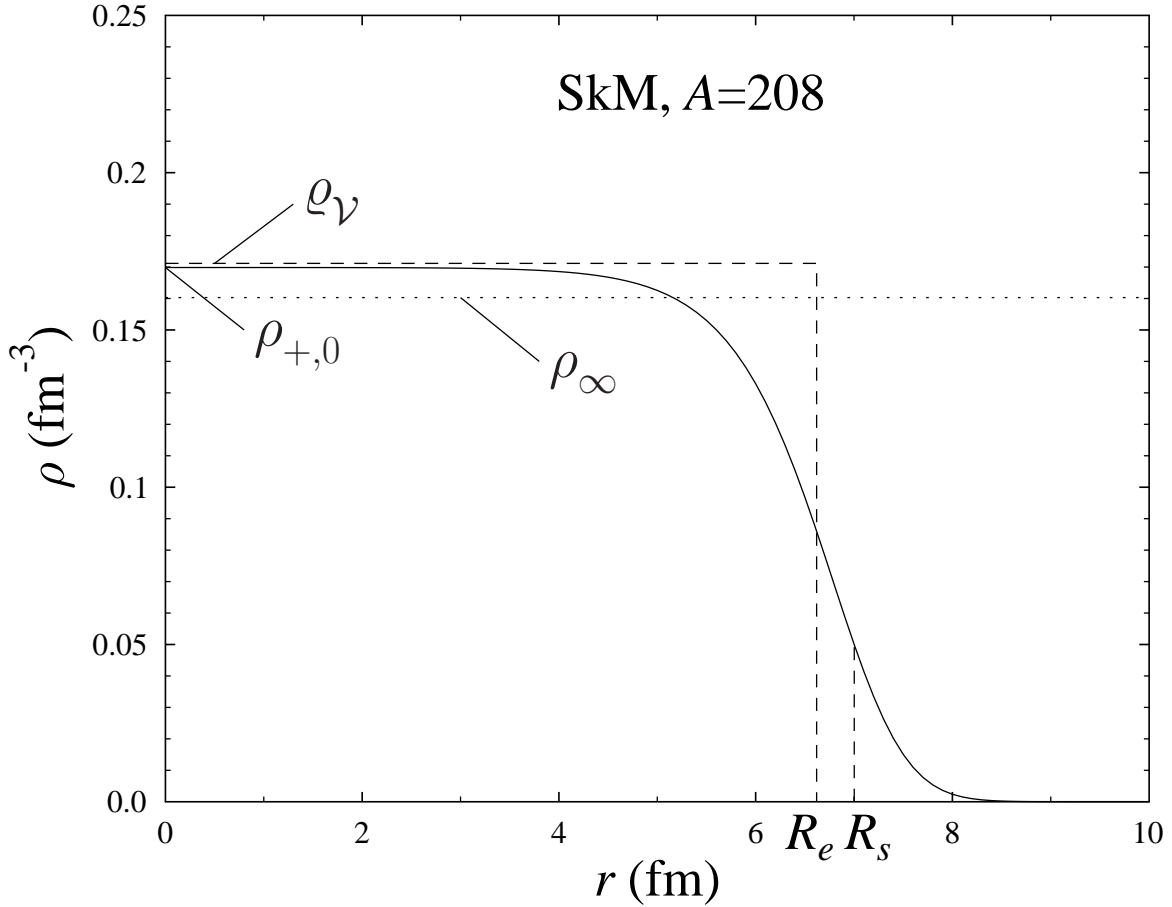}
\caption{Profile density $\protect\rho (r)$\ for $A=208$. Solid line shows
calculation for the actual droplet, dashed line corresponds to the equimolar
distribution dotted line is the particle density $\protect\rho _{\infty }$
in nuclear matter. Calculation was performed using SkM force. $R_{e}$\
denotes the equimolar radius, $R_{s}$ is the radius of surface tension.}
\label{fig3}
\end{figure}

\figurename\ \ref{fig3} illustrates the profile density of the droplet (solid line)
and its volume part (dashed line). One can see from this figure that the
density of nuclear matter, $\varrho _{\mathcal{V}}$, slightly differs from
that of the droplet bulk, $\rho _{+,0}$. Using SkM force for $A=208$ we
obtain slightly different values of $\varrho _{\mathcal{V}}=0.171$~fm$^{-3}$
and $\rho _{+,0}=0.170$~fm$^{-3}$. This difference will disappear for
incompressible liquid or in the planar limit. In ref.~\cite{Myer69} the
approximation $\rho _{+,0}=\varrho _{\mathcal{V}}$ was used when obtaining
the curvature correction to the surface tension. Since the correction for
curvature is calculated in the limit of semi-infinite matter, such
approximation will, obviously, give correct results. Note also that both
particle densities $\varrho _{\mathcal{V}}$\ and $\rho _{+,0}$\ exceed the
nuclear matter density $\rho _{\infty }$\ (dotted line in \figurename\ \ref{fig3}).
That is because the surface pressure, which influences the bulk properties,
leads to an increase in the nucleon density in center of the nucleus. 

Considering the arbitrary choice of the dividing surface we have determined
two radii, the equimolar dividing radius $R_{e}$ which corresponds to zero
surface density $\varrho _{\mathcal{S}}$\ and the radius of tension $R_{s}$
which corresponds to the minimum value of surface tension. From eqs.~(\ref%
{sigmaR}), (\ref{gammaR}) the values of these radii are given by 
\begin{equation}
R_{e}=\left( \frac{4\pi \varrho _{\mathcal{V}}}{3A}\right) ^{-1/3}\ ,\ \ \
R_{s}=\left( -\frac{2\pi \varrho _{\mathcal{V}}}{3A}\,\frac{\omega _{%
\mathcal{V}}}{\omega }\right) ^{-1/3}\ .  \label{Re}
\end{equation}%
Below we will assume that the physical (measurable) value of surface tension
is that taken at the equimolar dividing surface. Taking eq.~(\ref{p2}) for $%
R=R_{s}$, using eqs.~(\ref{p3}) and (\ref{sigmaeq}) and introducing small
value $\eta =R_{e}-R_{s}$, we obtain 
\begin{equation}
P=\frac{2\sigma _{\infty }}{R_{s}}\left( 1-\frac{2\xi }{R_{s}}+\mathcal{O}%
(R_{s}^{-2})\right) \ .  \label{p4}
\end{equation}%
Taking eq.~(\ref{p2}) for $R=R_{e}$ and eqs.~(\ref{sigmaeq}) we find 
\begin{equation}
P=\frac{2\sigma _{\infty }}{R_{s}}\left( 1-\frac{\xi +\eta }{R_{s}}+\mathcal{%
O}(R_{s}^{-2})\right) \ .  \label{p5}
\end{equation}%
We note, that one should make a difference between formal derivative $\sigma
^{\prime }[R]$\ in (\ref{p2}) and $\sigma ^{\prime }(R)$\ where the surface
tension is treated as a function of physical size. However, for the special
case of the equimolar dividing surface one can prove that $\sigma ^{\prime
}[R_{e}]=\sigma ^{\prime }(R_{e})$, see \cite{Wido82}. In particular, using
eq.~(\ref{sigmaeq}) one finds $\sigma ^{\prime }(R_{e})=\sigma _{\infty
}\left( 2\xi R_{e}^{-2}+\mathcal{O}\left( R_{e}^{-3}\right) \right) $\ and,
in contrast to $\sigma ^{\prime }[R_{s}]=0$, one has%\\
$\sigma ^{\prime}(R_{s})=\sigma _{\infty }\left( 2\xi R_{s}^{-2}+\mathcal{O}\left(
R_{s}^{-3}\right) \right) $. Comparing eqs.~(\ref{p4}) and (\ref{p5}) for $%
R_{s}\rightarrow \infty $, we obtain Tolman result \cite{tolm49} (see also 
\cite{Nijm92}) 
\begin{equation}
\xi =\lim_{A\rightarrow \infty }({R_{e}-R_{s})}\ .  \label{ksi}
\end{equation}%
This result leads to the important conclusions which were not mentioned in
previous studies of nuclear surface. First, one needs to define two
different radii: the equimolar radius, $R_{e}$, for the proper separation of
the surface energy from the total energy of nucleus and determination of the
droplet size, and the radius of tension, $R_{s}$, to determine the capillary
pressure. Second, to obtain the non-zero value of Tolman length, and,
consequently, the value of the curvature correction $\Delta \sigma _{\mathrm{%
curv}}\neq 0$, the droplet must have the finite diffuse surface layer.

%This result leads to the important conclusion that the curvature correction
%to the surface tension $\sigma $\ is derived by both\ the equimolar and the
%surface tension radii and disappears in the case of sharp surface. This fact
%was not taken into account in previous considerations of the curvature
%correction $\Delta \sigma _{\mathrm{curv}}$\ in nuclei, see, e.g.
%refs.~\cite{Myer69,mysw98,Stoc1973,podu03,Myer1973,tyap71,bera74,brguho84,gram83,far85,stbanisi88,duschvi93,moni95}.

The value of Tolman's length could be positive or negative. Positive value
of Tolman's length $\xi >0$ means $\sigma_{e}<\sigma_{\infty}$ (see eq.~(\ref%
{sigmaeq})) and negative one gives $\sigma_{e}>\sigma_{\infty}$ for curved
surface.

\section{Numerical results}

\label{sec:4} 
\begin{figure}
\includegraphics[width=0.95\columnwidth]{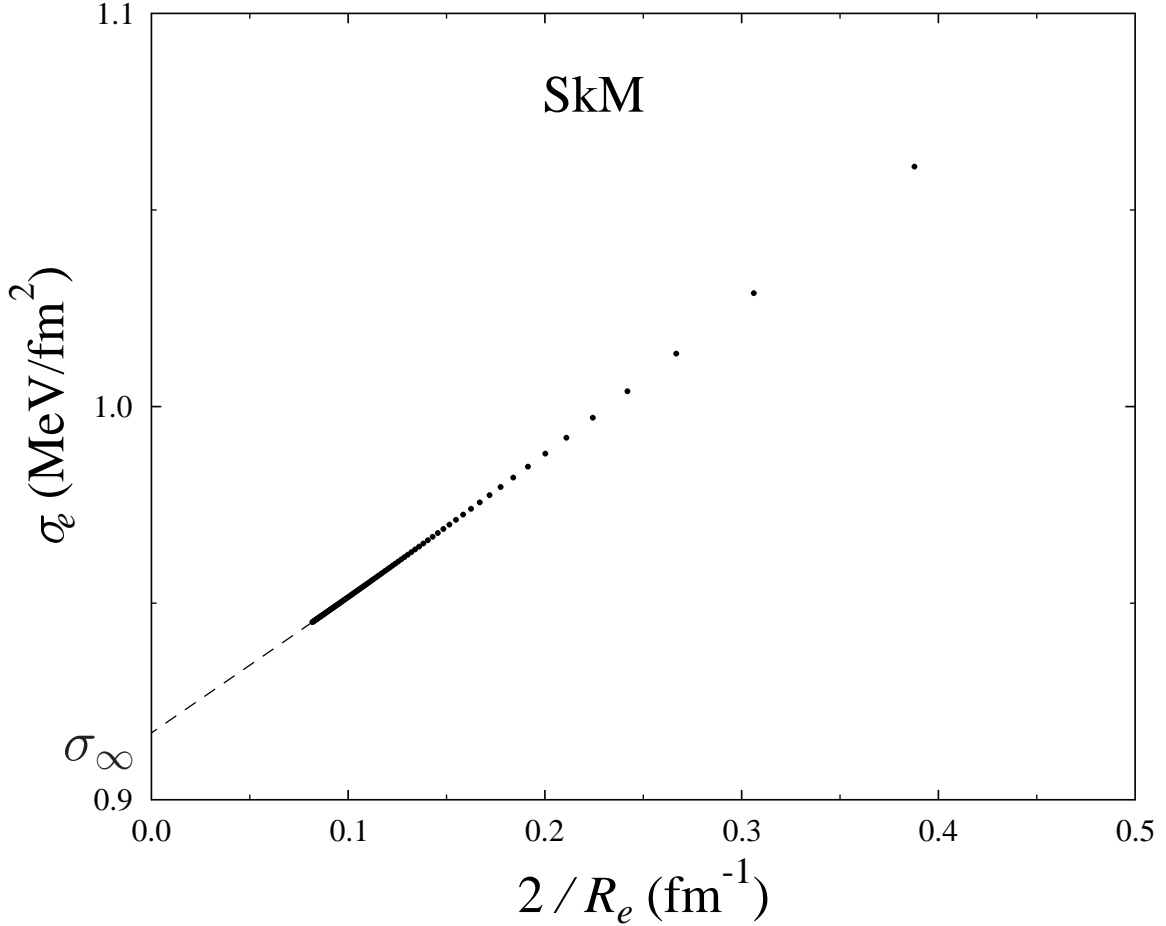}
\caption{Surface tension of the droplet versus the surface curvature for the
range of particle number $A=$~10$^{2}$\ $\div $ 10$^{4}$. Calculation was
performed using SkM force.}
\label{fig4}
\end{figure}

Since we consider a non-charged droplet (without Cou\-lomb),
the calculations is possible up to very high values of particle number $%
A\sim 10^{6}$. \figurename\ \ref{fig4} shows the result of calculation for tension $%
\sigma _{e}$ as a function of doubled droplet curvature $2/R_{e}$.
Calculation was carried out using SkM force. \figurename\ \ref{fig4} demonstrates
the negative value of $\xi $ for this calculation. An extrapolation of curve
in \figurename\ \ref{fig4} to zero curvature $2/R_{e}\rightarrow 0$ allows to derive
both the surface tension coefficient $\sigma _{\infty }=\sigma
_{e}(R_{e}\rightarrow \infty )$ in a planar geometry and the slope of curve
which gives the Tolman length $\xi $. The result of such kind of
extrapolation of $\sigma _{e}(R_{e})$\ is shown in \figurename\ \ref{fig4} by dashed
line.

We have determined the Tolman's length $\xi $ and the planar surface tension 
$\sigma _{\infty }$ for several parametrization of Skyrme interaction. For
this purpose we have fulfilled calculations up to particle number 10$^{6}$
and extrapolate them to zero curvature. Results are summarized in table~\ref%
{tab1}. To obtain the error of the extrapolation $2/R_{e}\rightarrow 0$\ we
estimated the magnitude of the higher order term $\sim R_{e}^{-2}$\ in (\ref%
{sigmaeq}). For the interval of particle numbers $A=$~10$^{4}$\ $\div $\ 10$%
^{6}$\ we gain the term of about $0.5R_{e}^{-2}$\ for the SkM interaction,
so one has here about of 10$^{-2}$\% contribution from this term to the
surface tension. One should expect the same accuracy for the extracted
values $\sigma _{\infty }$\ and $\xi $.

% Table 1
\begin{table}
\caption{
Values of Tolman's length $\xi$ and
planar surface tension $\sigma_{\infty}$ obtained for different
parameterizations of Skyrme forces.
}
\label{tab1}
~~~~~\begin{tabular}{lll}
\hline\noalign{\smallskip}
 Force\ \ \ \ \ \ \ \ \ \ \ \ \ \ \ \ &
 $\xi$ (fm)\ \ \ \ \ \ \ \ &
 $\sigma_{\infty}$ (MeV/fm$^{2}$) \\
\noalign{\smallskip}\hline\noalign{\smallskip}
SkM     & -0.36 & \ \ \ \ \ 0.92 \\
SIII    & -0.26 & \ \ \ \ \ 0.93 \\
SLy230b & -0.37 & \ \ \ \ \ 1.01 \\
T6      & -0.36 & \ \ \ \ \ 1.02 \\
\noalign{\smallskip}\hline
\end{tabular}
\end{table}

We can see from table~\ref{tab1} that Tolman's length $\xi $ is negative for
nuclear Fermi-liquid drop. This conclusion is also supported by the results
of ref.~\cite{Myer69}. The value of the Tolman length is only slightly
sensitive to the Skyrme force parametrization with the exception of old one
SIII.

The calculation of the curvature correction to the surface tension by use
the expansion around the plane surface (semi-infinite nuclear matter) with
respect to the surface curvature was introduced in \cite{Myer69} and it was
widely used for different types of nucleon-nucleon interactions including
the Skyrme-type interactions (see, for example, \cite{brguho84,CeVi93}.)
Comparing the values the surface, $a_{2}$, and curvature, $a_{3}$,
coefficients obtained for T6 and SIII forces in \cite{CeVi93} with the
analogous results for the same forces of table~\ref{tab1} by means of eqs.~(%
\ref{droplet-ksi}), (\ref{siginf}) (see the next section), one can see the
numerical coincidence of our results with that of \cite{CeVi93}. The main
reason of this coincidence, in our opinion, is the following. For the Gibbs
-- Tolman approach at the limit $A\rightarrow \infty $\ one has $\varrho _{%
\mathcal{V}}=\rho _{\infty }$\ and, consequently, the equimolar radius given
by eq. (\ref{Re}) and obtained from the condition $\varrho _{\mathcal{S}}=0$%
\ becomes equal to the equivalent "sharp" radius of the approach proposed in
Ref. \cite{Myer69}. In other words, the applicability of the Myers --
Swiatecky approach is realized in this case. The more detailed comparison is
quite difficult since the Gibbs -- Tolman approach does not rely \ on the
bulk asymptotics of the energy density functional.

In fact, the above comparison shows the equivalence of two approaches at
large masses. It is interesting to analyze the applicability of them for the
case of small mass numbers. Following \cite{brguho84,CeVi93} the
coefficients of mass formula are calculated using the leptodermous
approximation which requires the surface layer thickness to be small as
compared to the nuclear size given by the corresponding sharp radius.
According to Gibbs \cite{gibbs} the thermodynamical relation (\ref{capil})
remains exact up to the zero value of $R_{s}$, provided that the pressure is
calculated for the matter at the value of chemical potential of the actual
drop. Another conclusion concerning the Gibbs -- Tolman approach was made in 
\cite{Wido82}, namely, the low limit of $R_{s}$\ where the definition of the
surface tension make sense is about of $R_{s}\sim |\xi |$. In any case, even
though we will require the absolute value of the Tolman length to be small
as compared to $R_{s}$, the Gibbs -- Tolman definition of the surface
tension seems more preferable than that obtained using the leptodermous
approximation. The reason is that the estimated value for $|\xi |$\ (see
table~\ref{tab1}) is lower than the thickness, $t$, of the surface layer
(see \cite{CeVi93}, table 5).

\section{Link to the droplet model}

\label{droplet}

The Gibbs concept of dividing surface does not imply any specific energy
density functional and relies on the value of energy and the chemical
potential which are measurable quantities. It is possible to apply this
concept to the droplet model as well. We will consider non-charged ($N=Z$,
without Coulomb interaction) droplet at zero temperature and apply the same
procedure as described in previous sections to extract the value of Tolman
length. According to \cite{Myer69}, one can write free energy, $F$, and the
chemical potential, $\lambda $, of the nucleus having mass number $A$ as 
\begin{equation}
F=-a_{1}A+a_{2}A^{2/3}+\left( a_{3}-\frac{2a_{2}^{2}}{K}\right) A^{1/3}\ ,
\label{droplet-fin}
\end{equation}%
\begin{equation}
\lambda =-a_{1}+\frac{2}{3}a_{2}A^{-1/3}+\frac{1}{3}\left( a_{3}-\frac{%
2a_{2}^{2}}{K}\right) A^{-2/3}\ ,  \label{chem-fin}
\end{equation}%
where $a_{1}$, $a_{2}$ and $a_{3}$ are, respectively, the volume, the
surface and the curvature correction coefficients, $K$ is the
incompressibility coefficient. From eqs. (\ref{droplet-fin}) and (\ref%
{chem-fin}) one has the grand potential per particle $\omega =F/A-\lambda $
as
\begin{equation}
\omega =\frac{1}{3}a_{2}A^{-1/3}+\frac{2}{3}\left( a_{3}-\frac{2a_{2}^{2}}{K}%
\right) A^{-2/3}  \label{gppp-fin}
\end{equation}%
The equation of state for infinite nuclear matter in terms of droplet model
reads 
\begin{equation}
e=-a_{1}+\frac{1}{2}K\epsilon ^{2}  \label{droplet-inf}
\end{equation}%
for the free energy per particle, $e$, and 
\begin{equation}
l=-a_{1}+\frac{1}{6}K\epsilon \left( 9\epsilon -2\right)   \label{chem-inf}
\end{equation}%
for the chemical potential, $l$, beyond the equilibrium point. In eqs.~(\ref%
{droplet-inf}) and (\ref{chem-inf}), the dimensionless variable 
\begin{equation}
\epsilon =-\,\frac{1}{3}\,\frac{\rho -\rho _{\infty }}{\rho _{\infty }}
\end{equation}%
was introduced as the measure of difference between nuclear matter density $%
\rho $ and its equilibrium value $\rho _{\infty }$. Fixing the value of
particle density $\varrho _{\mathcal{V}}=\rho (\lambda )$ from the condition 
$l(\rho )=\lambda $, one obtains the volume part of grand potential per
particle 
\begin{equation}
\omega _{\mathcal{V}}=-\,\frac{2}{3}a_{2}A^{-1/3}-\,\frac{1}{3}\left( a_{3}-%
\frac{8a_{2}^{2}}{K}\right) A^{-2/3}+\mathcal{O}(A^{-1})\ ,  \label{gpppvol}
\end{equation}%
and, using also (\ref{gppp-fin}), the ratio $\omega /\omega _{\mathcal{V}}$ 
\begin{equation}
\frac{\omega }{\omega _{\mathcal{V}}}=-\,\frac{1}{2}-\,\frac{3}{4}\,\frac{%
a_{3}}{a_{2}}A^{-1/3}+\mathcal{O}(A^{-2/3})\ .  \label{gpprat}
\end{equation}%
As seen from eq. (\ref{gpprat}) \ the compression effect is canceled out from 
$\omega /\omega _{\mathcal{V}}$ up to the order of $A^{-1/3}.$ Using eqs.~(%
\ref{Re}) and (\ref{ksi}), one derives both radii 
\begin{equation}
R_{e}=r_{0}A^{1/3}\left[ 1-\frac{2a_{2}}{K}A^{-1/3}+\mathcal{O}(A^{-2/3})%
\right] \ ,  \label{droplet-Re}
\end{equation}%
\begin{equation}
R_{s}=r_{0}A^{1/3}\left[ 1+\left( \frac{a_{3}}{2a_{2}}-\frac{2a_{2}}{K}%
\right) A^{-1/3}+\mathcal{O}(A^{-2/3})\right]   \label{droplet-Rs}
\end{equation}%
and the Tolman length 
\begin{equation}
\xi =-\frac{a_{3}}{2a_{2}}\,r_{0}  \label{droplet-ksi}
\end{equation}%
where $r_{0}=(4\pi \rho _{\infty }/3)^{-1/3}$. Taking the eq.~(\ref{sigma})
at $R=R_{e}$, by the use of the eqs.~(\ref{gppp-fin}), (\ref{gpppvol}) and (%
\ref{droplet-Re}), the surface tension reads 
\begin{equation}
\sigma _{e}=\frac{(\omega -\omega _{\mathcal{V}})A}{4\pi R_{e}^{2}}=\frac{1}{%
4\pi r_{0}^{2}}\left( a_{2}+a_{3}A^{-1/3}+\mathcal{O}(A^{-2/3})\right) \ .
\label{droplet-sigma}
\end{equation}%
With Tolman length given by (\ref{droplet-ksi}) and relation 
\begin{equation}
\sigma _{\infty }=\frac{a_{2}}{4\pi r_{0}^{2}}  \label{siginf}
\end{equation}%
one can reduce eq.~(\ref{sigmaeq}) to (\ref{droplet-sigma}). As seen from
the above eqs.~(\ref{droplet-Re}), (\ref{droplet-Rs}) and (\ref{droplet-ksi}%
), both the equimolar, $R_{e}$, and the tension, $R_{s}$, radii include the
term of compression effect $(2a_{2}/K)A^{0}$, whereas the value of Tolman's
length $\xi $ of eq. (\ref{ksi}) reflects purely the effect of curvature of
dividing surface. Using the results presented in table~\ref{tab1}, one may
estimate the ratio of the curvature correction to the surface coefficient of
droplet model as $a_{3}/a_{2}\approx 0.63$ for the case of SkM
nucleon-nucleon interaction. This value of the ratio $a_{3}/a_{2}$ is
consistent with that of \cite{mysw98}.

\section{Conclusions}

\label{concl} Considering a small droplet with a finite diffuse layer, we
have introduced a formal dividing surface of radius $R$ which splits the
droplet onto volume and surface parts. The corresponding splitting was also
done for the free energy. Assuming that the dividing surface is located
close to the interface, we are then able to derive the volume pressure $P$
and the surface free energy $F_{\mathcal{S}}$. In general, the surface free
energy $F_{\mathcal{S}}$ includes the contributions from the surface tension 
$\sigma $ and from the binding energy of $A_{\mathcal{S}}$ particles within
the surface layer. The equimolar surface and the actual physical size of the
droplet was derived by the condition $\varrho_{\mathcal{S}}=0$.

In a small nucleus, the diffuse layer and the curved interface affect the
surface properties significantly. In agreement with Gibbs-Tolman concept 
\cite{tolm49,gibbs}, two different radii have to be introduced in this case.
The first radius, $R_{s}$, is the surface tension radius which provides the
minimum of the surface tension coefficient $\sigma $ and the satisfaction of
the Laplace relation (\ref{p3}) for capillary pressure. The another one, $%
R_{e}$, is the equimolar radius which corresponds to the equimolar dividing
surface and defines the physical size of the sharp surface droplet, i.e.,
the surface at which the surface tension is applied. The difference of both
radii ${R_{e}-R_{s}}$ in an asymptotic limit of large system $A\rightarrow
\infty $ derives the Tolman length $\xi $. That means that the presence of
curved surface is not sufficient for the calculation of the curvature
correction to the surface tension. The finite layer in the particle
distribution is required. %That means that the curvature
%correction to the surface tension depends on the finite diffuse layer in the
%particle distribution.
%This fact is usually ignored in the case of the
%phenomenological approaches founded on the polynomial fit in power of $%
%A^{1/3}$ to the nuclear masses and the fission barriers, see, e.g.
%ref.~\cite{moni95} where it was derived $\xi =0$. 
We point out that the Gibbs-Tolman theory allows to treat a liquid drop
within thermodynamics with minimum assumptions. Once the binding energy and
chemical potential\ of the nucleus are known its equimolar radius, surface
tension radius and surface energy can be evaluated by use of equation of
state of the infinite nuclear matter. In this sense, in contrast to the
"geometrical" definition of nuclear size \cite{Myer1973}, the Gibbs-Tolman
approach does not rely on details of the particle density profile. In
particular, the quantum oscillations of bulk density \cite{FaSt1985} do not
need to be smoothed to obtain the volume density $\varrho _{\mathcal{V}}$
which is different, in general, than the bulk density.

The sign and the magnitude of the Tolman length $\xi$ depend on the
interparticle interaction. We have shown that the Tolman length is negative
for the nuclear Fermi liquid drop. As a consequence of this the curvature
correction to the surface tension could lead to the hindrance of the yield
of light fragments at the nuclear multifragmentation in heavy ion collisions.

%\vspace*{-1.5ex}

\end{document}